\documentclass[a4paper,twocolumn]{article}
\usepackage{color}
\usepackage{setspace}

\usepackage{geometry}
\usepackage[figuresright]{rotating}
\usepackage{graphicx}
\usepackage{longtable}
\usepackage{array}

\usepackage{multirow}
\usepackage{makecell}

\usepackage{tensor}
\usepackage{mathtools}
\usepackage{amsmath}
\usepackage{mathrsfs}
\usepackage{upgreek}

\usepackage{appendix}

\usepackage{bm}

\usepackage[noblocks]{authblk}

\usepackage[numbers,sort&compress]{natbib}

\usepackage{CJK}
\usepackage{txfonts}
\usepackage{times}
\usepackage{ragged2e}
\usepackage{indentfirst}

\usepackage{verbatim}

\usepackage{ntheorem}

\renewcommand{\raggedright}{\leftskip=0pt \rightskip=0pt plus 0cm}

\title{Cryptanalysis and improvement of several quantum private comparison protocols}

\author{Zhao-Xu Ji$^{\dag}$, Pei-Ru Fan, Huan-Guo Zhang, Hou-Zhen Wang
\\
{\small Key Laboratory of Aerospace Information Security and Trusted
Computing, Ministry of Education, School of Cyber Science and
Engineering, Wuhan University, Wuhan 430072, China	\\
$^{\dag}$jizhaoxu@whu.edu.cn}}
\date{}

\begin{document}


\maketitle

\rmfamily
\fontsize{10}{13}\selectfont 


\section*{Abstract}

{\small

Recently, Wu et al. [Int. J. Theor. Phys. 58, 1854, (2019)]
found a serious information leakage problem in
Ye and Ji's quantum private comparison protocol
[Int. J. Theor. Phys. 56, 1517, (2017)], that is,
a malicious participant can steal another's secret data 
without being detected through an active attack means.
In this paper, we show that Wu et al.'s attack means is also effective for 
several other existing protocols, including the ones proposed by Ji et al. and Zha et al.
[Commun. Theor. Phys. 65, 711, (2016) and Int. J. Theor. Phys. 57, 3874, (2018)].
In addition, we propose a passive attack means,
which is different from Wu et al.'s active attack in that the malicious participant 
can easily steal another's secret data only by using his own secret data
after finishing the protocol.
Furthermore, we find that several other existing quantum private comparison protocols also 
have such an information leakage problem.
In response to the problem, we propose a simple solution,
which is more efficient than the ones proposed by Wu et al.,
because it does not consume additional classical and quantum resources.
We also make some comments on this problem.

\noindent
\textbf{Keywords}: quantum information security, 
quantum cryptography, quantum private comparison,
information leakage problem, passive attack

}


\section{Introduction}

Quantum cryptography is widely concerned because of its unconditional security
\cite{ZhangHG58112015,ZhangHG16102019,JiZX1862019}.
The difference between quantum cryptography and classical cryptography is that
the security of the former is based on some principles of quantum mechanics,
while the latter is based on some assumptions of computational complexity.
A fascinating feature of quantum cryptography is that it enables users to
detect whether there is an eavesdropper in quantum channels during communications,
which can not be done by classical cryptography \cite{ZhangHG16102019,JiZX1862019}. 
With the rapid development of quantum computers and quantum algorithms,
the security of classical cryptography has been severely challenged,
which makes the role of quantum cryptography in modern cryptography more and more important
\cite{ZhangHG16102019,JiZX1862019}.

Since the birth of quantum cryptography, 
quantum key distribution (QKD) has been one of the main research directions 
in quantum cryptography domain \cite{ZhangHG16102019}. Indeed, 
the first quantum cryptography protocol is the QKD protocol proposed by Bennett et al. in 1984, 
which is known as BB84 protocol. QKD aims to generate random shared keys 
between different users; combined with one-time pad encryption, 
it can provide unconditional security for users.
Moreover, the decoy photon technology derived from QKD has 
become one of the effective means for eavesdropping checking
\cite{JiZX098752019,YangYG4252009,LiuWJ3052013}.

Quantum private comparison (QPC), originated from the famous ``millionaires' problem"
\cite{YangYG4252009,LiuWJ3052013,YaoAC821982},
aims to judge whether the date of at least two
users who do not trust each other are the same or not while maintaining data privacy
using some quantum mechanics laws.
In fact, the comparison of the equality of data is widely used in real life,
including secret bidding and auctions, secret ballot elections,
e-commerce, and data mining \cite{ZhangHG16102019}.
One of the common applications is the identification of a system for users,
which aims to judge whether the users' secret information (e.g., password and fingerprint) 
is the same as the secret information stored in the system.
QPC can also solve the ``Tierc\'e problem'',
which is also known as the ``socialist millionaires' problem'' \cite{BoudotF1112001}.

After about ten years of development, QPC has attracted extensive attention 
in academia. Many protocols have been proposed based on 
different quantum states or different quantum technologies
\cite{JiZX72019,ChenXB28371010,JiZX34282019,JiZX1249112019,JiZX19112019,
LiuW51112012,LinS52112013,
LiJ5372014,SunZ5212013,XuGA10042012,
TsengHY1122012,HeGP9392018,LiYB5282013,LiC1852019,HuangW5692013,
LiL7512019,GuoFZ1282013,PanHM56102017,
XuL7332019,LiuB1672017,JiaHY5142012,JZX1672017,SongX72019,
AbulkasimH72019,HuangSL14112015,WangQL13112014,YeTY6092017}.
Unfortunately, information leakage often occurs;
many existing QPC protocols have been proved to be insecure
\cite{WuWQ5862019,JiS1022015,PanHM5772018,
LiuWJ6222014,WangC11042013,LiuXT8762013,ChangY3312016}.
Recently, Wu et al. \cite{WuWQ5862019} pointed out that 
there is a serious information leakage problem 
in Ye et al.'s QPC protocol \cite{YeTY5652017}; 
they showed that one participant in the protocol can 
steal another's secret information through an active attack means.
To solve this problem, they put forward two solutions: 
One is to use a QKD protocol to establish two new key sequences,
and use hash functions to complete a mutual authentication process;
the other is to use a QKD protocol to establish a new key sequence
and adopt unitary-operation-based symmetric encryption technology.
Although the two solutions ensure the security, however,
they both greatly reduce the efficiency of the protocol. 
On the one hand, both solutions use QKD to prepare additional keys,
which obviously increases resource consumption.
On the other hand, the hash functions and unitary operations
need additional quantum devices and technologies,
which greatly reduces the feasibility of the protocol.
After all, Ye et al.'s protocol does not use any other quantum technology
except for the necessary ones
such as the technology of preparing quantum states and quantum measurement.

In this paper, we will show that the active attack means proposed by Wu et al.
is also effective for the protocols presented in 
Refs.~\cite{JiZX6567112016,ZhangWW5352014,ZhaXW57122018}.
That is, these protocols are insecure under the attack.
However, we will propose a passive attack means to show
that a malicious participant can easily steal another's secret data
without using Wu et al.'s active attack means.
Specifically, after the end of the protocol,
the malicious participant can steal another's secret data 
only by using his own secret data.
Moreover, we will point out that the passive attack is effective not only
for the protocols presented in 
Refs.~\cite{YeTY5652017,JiZX6567112016,ZhangWW5352014,ZhaXW57122018},
but also for the protocols presented in Refs.~\cite{WangF59112016,ZhangWW5252013}.
Finally, we will propose a simple and effective solution to the information leakage problem
and make some comments. The rest of the paper is arranged as follows:
In Sec.~2, we review briefly the protocol proposed by Ji and Ye \cite{JiZX6567112016}.
In Sec.~3, we first take Ji and Ye's protocol as an example to show that
Wu et al.'s active attack is also effective to the protocols
presented in Refs.~\cite{JiZX6567112016,ZhangWW5352014,ZhaXW57122018},
and then we describe our passive attack means.
Sec.~5 introduces our solution to the information leakage problem
and gives our comments. Sec.~6 summarizes this paper.

\section{Review on Ji and Ye's protocol}

Let us review the QPC protocol proposed by Ji and Ye \cite{JiZX6567112016}.
Their protocol uses the highly entangled six-qubit genuine state
as information carriers, whose expression is given by
\begin{align}
\label{state_used}
\left|\varUpsilon\right\rangle=
\frac 1{\sqrt{32}}
&	\Big[\left|000000\right\rangle + \left|111111\right\rangle+\left|000011\right\rangle		\notag\\
+&\phantom{i}\left|111100\right\rangle + \left|000101\right\rangle + \left|111010\right\rangle  \notag\\
+&\phantom{i}\left|000110\right\rangle + \left|111001\right\rangle + \left|001001\right\rangle	\notag\\
+&\phantom{i}\left|110110\right\rangle + \left|001111\right\rangle + \left|110000\right\rangle	\notag\\
+&\phantom{i}\left|010001\right\rangle + \left|101110\right\rangle + \left|010010\right\rangle  \notag\\
+&\phantom{i}\left|101101\right\rangle + \left|011000\right\rangle + \left|100111\right\rangle 	\notag\\
+&\phantom{i}\left|011101\right\rangle + \left|100010\right\rangle - (\left|010100\right\rangle  	\notag\\
+&\phantom{i}\left|101011\right\rangle + \left|010111\right\rangle + \left|101000\right\rangle	\notag\\
+&\phantom{i}\left|011011\right\rangle + \left|100100\right\rangle + \left|001010\right\rangle  \notag\\
+&\phantom{i}\left|110101\right\rangle + \left|001100\right\rangle + \left|110011\right\rangle 	\notag\\
+&\phantom{i}\left|011110\right\rangle + \left|100001\right\rangle) \Big],
\end{align}
which is rewritten as
\begin{align}
\left|\varUpsilon\right\rangle &=
\frac 1 4 \Big[  
\Big(      \left|0000\right\rangle - \left|0101\right\rangle - 
\left|1010\right\rangle + \left|1111\right\rangle   \Big)
\otimes	\left|\phi^{+}\right\rangle
\notag\\
& +	\Big(     \left|0001\right\rangle + \left|0100\right\rangle + 
          \left|1011\right\rangle + \left|1110\right\rangle    \Big)
\otimes	\left|\psi^{+}\right\rangle
\notag\\
& + \Big(    \left|0110\right\rangle - \left|0011\right\rangle - 
         \left|1001\right\rangle + \left|1100\right\rangle    \Big)
\otimes	\left|\phi^{-}\right\rangle   							
\notag\\
& + \Big(     \left|0010\right\rangle + \left|0111\right\rangle - 
          \left|1000\right\rangle - \left|1101\right\rangle    \Big)
\otimes	\left|\psi^{-}\right\rangle  \Big],
\end{align}
where
\begin{equation}
\begin{split}
\left|\phi^{\pm}\right\rangle = \frac 1{\sqrt{2}}
\Big(   \left|00\right\rangle \pm \left|11\right\rangle  \Big),\quad
\left|\psi^{\pm}\right\rangle = \frac 1{\sqrt{2}}
\Big(   \left|01\right\rangle \pm \left|10\right\rangle  \Big)	,
\end{split}
\end{equation}
are four Bell states.
The prerequisites of the protocol are:

\begin{enumerate}

\item Suppose that Alice and Bob have the secret data $X$ and $Y$ respectively,
and that the binary representations of
$X$ and $Y$ are $\left(  x_{1},x_{2},\ldots,x_{N} \right)$ and
$\left(  y_{1}, y_{2},\ldots,y_{N}  \right)$ respectively, where
$ x_{j}, y_{j} \in \{ 0, 1 \} \phantom{1} \forall j \in \{ 1,2,\ldots,N \}$,
hence $X=\sum_{j=1}^{N} x_j2^{j-1},Y=\sum_{j=1}^{N} y_j2^{j-1}$.

\item Alice (Bob) divides the binary representation of $ X(Y) $ into $\lceil N/2 \rceil$ groups:
\begin{equation}
G_A^1,G_A^2,\ldots,G_A^{ \lceil \frac N 2 \rceil}
(G_B^1,G_B^2,\ldots,G_B^{\lceil \frac N 2 \rceil}).
\end{equation}
Each group $ G_A^i (G_B^i) $ includes two bits,
where $i = 1,2,\ldots,\lceil N/2 \rceil$ throughout this protocol.
If $ N $ mod  2 = 1, Alice (Bob) adds one 0 into the last group
$ G_A^{ \lceil N/2 \rceil} ( G_B^{\lceil N/2 \rceil} ) $.

\item Alice and Bob generate the shared key sequences
$\{K_{A}^1, K_{A}^2,\ldots, K_{A}^{\lceil N/2  \rceil}\}$
and $\{K_{B}^1, K_{B}^2,\ldots, K_{B}^{\lceil N/2  \rceil}\}$
through a QKD protocol, where $ K_{A}^i, K_{B}^i \in \{00,01,10,11\}$. 
Similarly, Alice(Bob) and TP generate the shared key sequence
$\{K_{AC}^1$, $K_{AC}^2$,$\ldots$, $K_{AC}^{\lceil N/2 \rceil}\}$
($\{K_{BC}^1$, $K_{BC}^2$,$\ldots$, $K_{BC}^{\lceil N/2 \rceil}\}$),
where $ K_{AC}^i, K_{BC}^i \in \{00,01,10,11\}$.

\item Alice, Bob and TP agree on the following coding rules:
$\left|0\right\rangle \leftrightarrow 0, \left|1\right\rangle \leftrightarrow 1,
\left|\phi^+\right\rangle \leftrightarrow 00, \left|\phi^-\right\rangle \leftrightarrow 11,
\left|\psi^+\right\rangle \leftrightarrow 01$, and $\left|\psi^-\right\rangle \leftrightarrow 10$.

\end{enumerate}

The steps of the protocol are as follows:

\begin{enumerate}

\item TP prepares $ \lceil N/2 \rceil $ copies of the highly entangled six-qubit genuine state
$\left|\varUpsilon\right\rangle$, and marks them by
\begin{align}
&\left|\varUpsilon(p_1^1,p_1^2,p_1^3,p_1^4,p_1^5,p_1^6)\right\rangle,
\left|\varUpsilon(p_2^1,p_2^2,p_2^3,p_2^4,p_2^5,p_2^6)\right\rangle,\notag\\
&\ldots,
\left|\varUpsilon(p_{\lceil N/2 \rceil}^1,p_{\lceil N/2 \rceil}^2,
p_{\lceil N/2 \rceil}^3,p_{\lceil N/2 \rceil}^4,
p_{\lceil N/2 \rceil}^5,p_{\lceil N/2 \rceil}^6)\right\rangle,
\end{align}
in turn to generate an ordered sequence, where the subscripts $1,2,\ldots,\lceil N/2 \rceil$
denote the order of the highly entangled six-qubit genuine states in the sequence, 
and the superscripts 1,2,3,4,5,6 denote six particles in one state.
Then TP takes the first two particles out from
$\left|\varUpsilon(p_i^1,p_i^2,p_i^3,p_i^4,p_i^5,p_i^6)\right\rangle$ 
to construct the new sequence
\begin{align}
p_1^1,p_1^2,p_2^1,p_2^2,\ldots,p_{\lceil N/2 \rceil}^1,p_{\lceil N/2 \rceil}^2,
\end{align}
and denotes it as $S_A$. Similarly, he takes out the third and fourth particles
to construct another new sequence
\begin{align}
p_1^3,p_1^4,p_2^3,p_2^4,\ldots,p_{\lceil N/2 \rceil}^3,p_{\lceil N/2 \rceil}^4,
\end{align}
and denotes it as $S_B$. The remaining particles construct another new sequence
\begin{align}
p_1^5,p_1^6,p_2^5,p_2^6,\ldots,p_{\lceil N/2 \rceil}^5,p_{\lceil N/2 \rceil}^6,
\end{align}
denoted as $S_C$.

\item TP prepares two sets of decoy photons in which each decoy photon 
is chosen randomly from the single-particle states
$\left|0\right\rangle, \left|1\right\rangle, \left|+\right\rangle, \left|-\right\rangle$ 
($ \left|\pm\right\rangle $ = $ 1/\sqrt{2}\left( \left|0\right\rangle\pm\left|1\right\rangle \right) $).
Then he inserts randomly the two sets of decoy photons into $ S_A $ and $ S_B $, respectively,
and records the insertion positions.
Finally, he denotes the two new generated sequences as $ S_A^{*} $ and $ S_B^{*} $,
and sends them to Alice and Bob, respectively.

\item After receiving $ S_A^{*} $ and $ S_B^{*} $, TP and Alice(Bob)
use the decoy photons in $S_A^{*}$ and $S_B^{*}$ to 
judge whether eavesdroppers exist in quantum channels.
The error rate exceeding the predetermined threshold will lead to the termination and
restart of the protocol, otherwise the protocol proceeds to the next step.

\item	Alice(Bob) measures the two particles marked by
$p_i^1,p_i^2$ ($p_i^3,p_i^4$) in $ S_A  (S_B) $ with $Z$ basis 
($\{\left|0\right\rangle,\left|1\right\rangle\}$),
and denotes the binary numbers corresponding to 
the measurement results as
$ M_A^i (M_B^i) $. Then, Alice(Bob) calculates 
$G_A^i \oplus M_A^i \oplus K_{AC}^i \oplus K_A^i$ 
($G_B^i \oplus M_B^i \oplus K_{BC}^i \oplus  K_B^i$),
and marks the calculation results by $R_A^i (R_B^i )$.
Finally, Alice(Bob) announces $ R_A^i (R_B^i) $ to TP.

\item After receiving $ R_A^i (R_B^i) $, TP performs Bell
measurements on the particles marked by $p_i^5,p_i^6$,
and marks the binary numbers corresponding to
the measurement results by $M_C^i$. Then, he calculates
$R_A^i \oplus R_B^i \oplus K_{AC}^i \oplus K_{BC}^i \oplus M_C^i$,
and marks the calculation results by $R_i$.
Finally, he announces $R_i$ to Alice and Bob.

\item After receiving $R_i$, Alice and Bob calculate 
$R_i \oplus K_A^i \oplus  K_B^i$, respectively, 
and mark the calculation results by $R_i^{\prime}$.
If $ R_i^{\prime}  =  00 $ (i.e., each classical bits in $ R_i^{\prime}$ is 0),
they conclude that their data $X$ and $Y$ are the same.
Otherwise, they conclude that $X$ and $Y$ are different
and stop the comparison.

\end{enumerate}

\section{Information leakage problem}

In this section, we will show that the protocol is insecure under Wu et al.'s active attack means:
a malicious participant can steal the secret information of another by forging identities. 
We will then propose a passive attack means by which the malicious participant 
can also steal the secret information of another.

\subsection{Information leakage under Wu et al.'s active attack}


Let us now show how a malicious participant steal another's secret information
by using Wu et al.'s active attack.
Without losing generality, we assume that Bob is malicious.
He can steal Alice's secret data through the following steps:

\begin{enumerate}

\item In the second step of Ji and Ye's protocol,
when TP sends the particle sequence $ S_A^{*} $ to Alice,
Bob intercepts all the particles in the sequence,
and then he pretends to be Alice and tells TP that he has received all the particles.

\item Bob continues to pretends to be Alice and completes eavesdropping checking with TP.
Then he performs single-particle measurements on the particles marked by $p_i^1,p_i^2$ in $S_A$,
and denotes the binary numbers corresponding to the measurement results as
$M_{AB}^i$. Finally, TP denotes the particle sequence after measurements as $S_A^{1}$.

\item Similar to the second step of Ji and Ye's protocol,
Bob prepares a set of decoy photons,
and then inserts randomly them into $S_A^{1}$.
The new generated sequence is denoted as $S_A^{1*}$.
Finally, Bob pretends to be TP and send $S_A^{1*}$ to Alice.

\item After confirming that Alice has received $S_A^{1*}$,
Bob continues to pretends to be TP and completes eavesdropping checking with Alice.
If there is no eavesdropping, according to the protocol procedures,
Alice measures each particle in $S_A^{1}$ with $Z$ basis,
and denotes the binary numbers corresponding to
the measurement results as $M_A^i$
(obviously, $M_A^i$ is the same as $M_{AB}^i$, i.e., $M_A^i=M_{AB}^i$).
Then she calculates $G_A^i \oplus M_A^i \oplus K_{AC}^i \oplus K_A^i$,
and marks the calculation results by $R_A^i$.
Finally, Alice announces $R_A^i$ to TP.
Similarly, Bob announces $R_B^i$ to TP after completing measurements and calculations 
in accordance with the protocol procedures.

\item According to the protocol procedures,
TP completes measurements, calculations and publishes $R_i$ to 
Alice and Bob. After receiving $R_i$, Bob can calculate
\begin{align}
& R_i \oplus K_{BC}^i \oplus M_C^i \oplus R_B^i \oplus K_A^i \oplus M_{AB}^i \notag\\
= & ( R_A^i \oplus R_B^i \oplus K_{AC}^i \oplus K_{BC}^i \oplus M_C^i ) 
	\oplus K_{BC}^i \oplus M_C^i \oplus R_B^i \notag\\
	& \oplus K_A^i \oplus M_{AB}^i \notag\\
= & R_A^i \oplus K_{AC}^i \oplus K_A^i \oplus M_{AB}^i \notag\\
= & (G_A^i \oplus M_A^i \oplus K_{AC}^i \oplus K_A^i) 
	\oplus K_{AC}^i \oplus K_A^i \oplus M_{AB}^i \notag\\
= & G_A^i.
\end{align}
Note here that $M_A^i=M_{AB}^i$, and Bob can deduce $M_C^i$
from Eq.~\ref{state_used} based on $M_{AB}^i$ and $M_{B}^i$.
From the above equation, Bob can obtain $G_A^i$ through the calculation,
thus he can deduce Alice's secret data $X$.

\end{enumerate}

We have shown that Wu et al.'s active attack is also effective for Ji and Ye's protocol, 
that is, their protocol will leak information under Wu's active attack.
In addition, we find that the protocols presented in 
Refs.~\cite{JiZX6567112016,ZhangWW5352014,ZhaXW57122018}
also have such an information leakage problem,
because the process of these protocols is similar to that of Ji and Ye's protocol.

In what follows, we will present a passive attack means, by which
we will show that a malicious participant can easily steal the secret data 
of another based on his own secret data after the end of the protocol, 
instead of using Wu et al.'s active attack means.

\subsection{Information leakage under the proposed passive attack}


At the end of the protocol, both Alice and Bob obtain 
$G_A^i \oplus G_B^i$ (i.e., $R_i^{\prime}$), that is,
\begin{align}
R_i^{\prime} &  =   R_i  \oplus  K_A^i \oplus  K_B^i 		\notag\\
&	= ( R_A^i \oplus R_B^i \oplus K_{AC}^i \oplus K_{BC}^i \oplus M_C^i )
		\oplus ( K_A^i \oplus  K_B^i )		\notag\\
&	= \big[( G_A^i \oplus M_A^i \oplus K_{AC}^i \oplus K_A^i )
		\oplus (  G_B^i \oplus M_B^i \oplus K_{BC}^i \oplus  K_B^i )		 \notag\\
&	\quad	\oplus K_{AC}^i \oplus K_{BC}^i \oplus M_C^i \big]
		\oplus ( K_A^i \oplus  K_B^i )      				\notag\\
&	= ( G_A^i \oplus  G_B^i )  \oplus (  M_A^i \oplus  M_B^i \oplus M_C^i )	  \notag\\
&	= G_A^i \oplus  G_B^i.
\end{align}
In this case, Alice and Bob can easily steal each other's data.
Specifically, Alice(Bob) can calculate $R_i^{\prime} \oplus G_A^i (R_i^{\prime} \oplus G_B^i)$,
thus she(he) can get $G_B^i (G_A^i)$, that is,
 $R_i^{\prime} \oplus G_A^i = (G_A^i \oplus G_B^i) \oplus G_A^i = G_B^i$
[$R_i^{\prime} \oplus G_B^i = (G_A^i \oplus G_B^i) \oplus G_B^i = G_A^i$].
In fact, for a cryptography protocol, the process, prerequisites,
and coding rules of the protocol are all public,
except that the keys generated in the protocol is confidential.
Therefore, Alice and Bob, as participants,
will surely know that the final comparison result is $G_A^i \oplus G_B^i$.

We find that the protocols in Refs. \cite{YeTY5652017,WangF59112016,
ZhangWW5252013,ZhaXW57122018,ZhangWW5352014}
also have such information leakage problem.
In these protocols, both Alice and Bob obtain $G_A^i \oplus G_B^i$ 
at the end of the protocol, thus they can easily know each other's data.

\section{New solution to the information leakage problem}

We have proposed a passive attack means, 
and described the information leakage problem of several QPC protocols under this attack. 
Indeed, the information leakage problem is the same as that
under Wu et al.'s active attack, i.e., 
two participants can steal each other's secret data.
To solve this problem, Wu et al. put forward two solutions, 
which is mentioned in the introduction.
In what follows, we will propose a new solution to 
the information leakage problem,
and then we will briefly compare our solution with that of Wu et al.
We will finally make some relevant comments.

\subsection{The proposed solution}

Let us now describe our solution.
For simplicity and clarity, we change directly the steps 5 and 6 
of Ji and Ye's protocol as follows (the first four steps of the protocol remain unchanged):

\begin{enumerate}

\item[5] After receiving $ R_A^i (R_B^i) $, TP performs Bell
measurements on the particles marked by $p_i^5,p_i^6$,
and marks the binary numbers corresponding to
the measurement results by $M_C^i$. Subsequently, TP calculates
$R_A^i \oplus R_B^i \oplus K_{AC}^i \oplus K_{BC}^i \oplus M_C^i$,
and marks the calculation results by $a_i^1a_i^2$
(note that each calculation result is a binary number containing two bits, i.e.,
$a_i^1a_i^2 \in \{00,01,10,11\}$). Then, TP calculates
\begin{align}
\sum_{i=1}^{\lceil N/2 \rceil}	\sum_{j=1}^2 a_i^j,
\end{align}
and marks the calculation result by $S$.
Finally, he announces $S$ to Alice and Bob.

\item[6] After receiving $S$, Alice and Bob calculate 
$K_A^i \oplus  K_B^i$, respectively,
and marks the calculation results by $b_i^1b_i^2$.
Then, they calculate
\begin{align}
\sum_{i=1}^{\lceil N/2 \rceil}	\sum_{j=1}^2 	b_i^j,
\end{align}
and marks the calculation result by $S^{\prime}$.
Finally, they calculate $S-S^{\prime}$.
If $S-S^{\prime} = 0$,
they can conclude that their data $X$ and $Y$ are the same. 
Otherwise, they conclude that $X$ and $Y$ are different.

\end{enumerate}

The correctness of our solution is easy to verify.
In Step 5, TP calculates 
$R_A^i \oplus R_B^i \oplus K_{AC}^i \oplus K_{BC}^i \oplus M_C^i$,
hence we get
\begin{align}
& R_A^i \oplus R_B^i \oplus K_{AC}^i \oplus K_{BC}^i \oplus M_C^i 		\notag\\
= &	( G_A^i \oplus M_A^i \oplus K_{AC}^i \oplus K_A^i )
		\oplus (G_B^i \oplus M_B^i \oplus K_{BC}^i \oplus  K_B^i )		 \notag\\
&	\oplus K_{AC}^i \oplus K_{BC}^i \oplus M_C^i \notag\\
= &	G_A^i \oplus  G_B^i \oplus K_A^i \oplus K_B^i.
\end{align}
Obviously, $S = \sum_{i=1}^{\lceil N/2 \rceil}	\sum_{j=1}^2 	b_i^j$ (i.e., $S=S^{\prime})$
if and only if $G_A^i = G_B^i$. Otherwise, $S \ne S^{\prime}$.
Note here that $K_A^i$ and $K_B^i$ are random keys generated by QKD,
thus $K_A^i$ and $K_B^i$ are not all the same
(the probability that they are all the same can be ignored because it is very small).

Similar improvements can be made to the protocols presented in
Refs.~\cite{YeTY5652017,WangF59112016,
ZhangWW5252013,ZhaXW57122018,ZhangWW5352014}.
For simplicity, we would not like to review these protocols and describe their amendments.


\subsection{Comparison}

Let us make a brief comparison between our solution and the ones proposed by Wu et al. 
In our solution, we only change slightly the algorithm  
without using any additional quantum technology and resources. 
On the contrary, both the solutions proposed by Wu et al. need to 
consume additional quantum technology and resources, 
which has been mentioned in the introduction.
We show these differences in Table~\ref{comparison}.

\begin{table}[h]
\setlength{\tabcolsep}{2.5pt}

\caption{Comparison with Wu et al.'s solutions.}
\label{comparison}
\centering
\begin{tabular}{cccc}

\hline\noalign{\smallskip}

 & \makecell{Wu et al.'s \\ solution 1} & \makecell{Wu et al.'s \\ solution 2} 
& \makecell{Our \\ solution} \\

\noalign{\smallskip}\hline\noalign{\smallskip}

\makecell{additional  keys} & $\surd$ & $\surd$ & $\times$ 	\\

\noalign{\smallskip}\noalign{\smallskip}

\makecell{hash functions} & $\surd$ & $\times$ & $\times$ \\

\noalign{\smallskip}\noalign{\smallskip}

\makecell{unitary operations} & $\times$ & $\surd$ & $\times$  \\

\noalign{\smallskip}\hline

\end{tabular}
\end{table}

\subsection{Comment}


In fact, in classical private comparison protocols, 
Alice and Bob's data are required to be confidential, 
but there is no requirement on whether the final comparison results are public
\cite{BoudotF1112001}.
Therefore, QPC, as the generalization of 
classical private comparison protocols in quantum mechanics, 
does not need to add such a privacy requirement.
After all, adding this requirement will inevitably make the protocol 
more complex and increase resource consumption
(e.g., consuming more expensive quantum devices).
At present, most QPC protocols allow the third party to publish the final comparison result
(i.e., the comparison result is public).
Of course, if there is such a requirement in reality (i.e., keeping the comparison result private),
one can design a protocol in a similar way according to our solution.

\section{Conclusion}

We have shown that several QPC protocols have the same
information leakage problem under Wu et al.'s active attack.
We have proposed a passive attack means, and shown that 
several QPC protocols are insecure under this attack:
a malicious participant can easily steal another's secret data after the end of the protocol.
We have proposed a simple and effective solution to this problem,
which is more efficient than the ones proposed by Wu et al.
We have also made some comments on this problem.
We believe that our solution and comments are 
constructive to the design of a QPC protocol.

\section*{Acknowledgments}

This work is supported by the State Key Program of
National Natural Science Foundation of China under Grant 61332019,
the Major State Basic Research Development Program of China (973 Program)
under Grant 2014CB340601, the National Science Foundation of China
under Grant 61202386 and Grant 61402339,
and the National Cryptography Development Fund of
China under Grant MMJJ201701304.


\end{document}